\documentclass[11pt]{article}

\usepackage{amssymb}
\usepackage{amsmath}
\usepackage{amsthm}
\usepackage{cite}
\usepackage{xspace}
\usepackage{hyperref}
\usepackage{url}
\usepackage{color}

\usepackage{subfigure}
\usepackage{psfig}
\usepackage{epsfig}
\usepackage{psfrag}       %latex text in .eps files
\usepackage[noend,ruled,linesnumbered]{algorithm2e}

\usepackage{listings}

\newtheorem{theorem}{Theorem}

\newtheorem{definition}[theorem]{Definition}

\newcommand{\parhead}[1]{{\textbf{#1.}\xspace}}

\newcommand{\RNs}{{Radio Networks}\xspace}

\newcommand{\MAN}{{Mobile Ad-hoc Network}\xspace}
\newcommand{\MANs}{{Mobile Ad-hoc Networks}\xspace}

\newcommand{\Dis}{{Dissemination}\xspace}
\newcommand{\OR}{{online route}\xspace}

\newcommand{\mess}{{information}\xspace}
\newcommand{\LA}{{locally adaptive}\xspace}

\begin{document}

\title{%Time Complexity of\\
Opportunistic Information Dissemination in\\
\MANs:\\
adaptiveness vs. obliviousness and\\
randomization vs. determinism
%\thanks{This research was supported in part by  }
%\thanks{We thank Seth Gilbert for triggering the development of this work asking a question at DISC 2010.}
}
\author{
   Mart\'in~Farach-Colton\thanks{Department of Computer Science, Rutgers University, Piscataway, NJ, USA. \href{mailto:farach@cs.rutgers.edu}{farach@cs.rutgers.edu}}
   \and
   Antonio~Fern\'andez~Anta\thanks{Institute IMDEA Networks, Madrid, Spain. \href{mailto:antonio.fernandez@imdea.org}{antonio.fernandez@imdea.org}}
   \and
   Alessia~Milani\thanks{LABRI, UniversitŽ Bordeaux 1, ENSEIRB, Talence, France. \href{mailto:milani@labri.fr}{milani@labri.fr}}
   \and
   Miguel~A.~Mosteiro\thanks{Department of Computer Science, Rutgers University, Piscataway, NJ, USA. \href{mailto:mosteiro@cs.rutgers.edu}{mosteiro@cs.rutgers.edu}} \thanks{LADyR, GSyC, Universidad Rey Juan Carlos, M\'ostoles, Madrid, Spain.}
   \and
   Shmuel~Zaks\thanks{Department of Computer Science, Technion - Israel Institute of Technology, Haifa, Israel. \href{mailto:zaks@cs.technion.ac.il}{zaks@cs.technion.ac.il}}
}

\date{}

\maketitle              % typeset the title of the contribution

%%%%%%%%%%%%%%%%%%%%%%%%%%%%%%%%%%%%%%%%%%%%%%%%%%%%%%%

\begin{abstract}
In this paper the problem of information dissemination in \MANs (MANET) is studied. The problem is to disseminate a piece of information, initially held by a distinguished source node, to all nodes in a set defined by some predicate.
%This problem abstracts many fundamental problems in MANETs, such as broadcast or gossiping.
We use a model of MANETs that is well suited for dynamic networks and opportunistic communication. In this model  nodes are placed in a plane, in which they can move with bounded speed, and communication between nodes occurs over a collision-prone single channel. In this setup informed and uninformed nodes can be disconnected for some time (bounded by a parameter $\alpha$), but eventually some uninformed node must become neighbor of an informed node and remain so for some
time (bounded by a parameter $\beta$). In addition, nodes can start at different times, and they can crash and recover.
Under the above framework, we show negative and positive results for different types of randomized protocols, and we put those results in perspective with respect to previous deterministic results.
%
%\textcolor{blue}{I would remove the rest.}
%Among other results, we show that even using some adaptiveness protocols require expected time $\Omega(\alpha n + n^2/\log n)$
%to solve the problem in a network with $n$ nodes, even if all nodes start simultaneously and do not fail. Then, we show that a matching upper bound can be achieved with high probability by a simple protocol. This upper bound is at least logarithmically asymptotically more efficient than any deterministic protocol.
%%The model includes parameters that characterize the connectivity and stability properties under which nodes move, fail or are activated. These parameters characterize  \emph{any} model of dynamic networks, and affect the progress that a protocol \emph{may} achieve in solving basic tasks.
\end{abstract}

%\newpage
%-------------------------------------------------------------------------------------------------------
%\setcounter{page}{1}
\section{Introduction: Background and Preliminaries}

A MANET is a network model defined by a set of processing nodes that move in an environment that lacks any form of communication infrastructure. Although each node is equipped with a radio to allow ad-hoc communication, node mobility and unreliability yield a dynamically changing network topology, which in turn makes the dissemination of information through the network a challenging task.
Should those changes be arbitrary, the most basic problems would not be solvable.
But, on the opposite end, strong stability assumptions - such as full connectivity - are frequently unrealistic.
To cope with such challenge, models such as \emph{delay-tolerant networks}~\cite{F:delaytolerant} and \emph{opportunistic networking}~\cite{PPC:opportunistic,FMMZ:geocast} have been recently proposed and studied.
A common theme underlying these models is a relaxed connectivity assumption, where every pair of nodes may communicate, but the path to achieve communication may be fragmented along a period of time. Under such models, it is said that the network allows \emph{opportunistic communication}.

In this paper we revisit a class of MANETs
that is well suited for opportunistic communication. Specifically, the model includes parameters that mainly characterize the connectivity and the stability properties of the network, provided that nodes move, may fail and may be activated at any time. These parameters characterize  \emph{any} model of dynamic networks, and affect the progress that a protocol \emph{may} achieve in solving basic tasks. 
In particular, we consider the problem of information dissemination.

%\marginpar{\textcolor{red}{Text removed here}}
%i.e.;  we want to disseminate a piece of information, initially held by a distinguished source node, to all nodes of a given set in the network. The dissemination problem abstracts many fundamental problems in MANETs such as broadcast, routing, etc.

%%%%%%%%%%%%%%%%%%%%%%%%%%%%%%%%%%%%%%%%%%%%

%\subsection{The \Dis Problem}
\label{section:prelim}

\parhead{The Problem}
We study the problem of disseminating a piece of information, initially held by a distinguished source node, to all nodes of a given set in the network.
%Briefly, \Dis is solved when a piece of information, initially held by a source node, is received by all nodes of a given set in the network.
Formally,

\begin{definition}
\label{def:geocast}
Given a MANET formed by a set $V$ of $~n$ nodes, let $\mathcal{P}$ be a predicate on $V$ and $s \in V$ a node that holds a piece of information $I$ at time $t_1$ ($s$ is the source of dissemination).
The \emph{\Dis} problem consists of distributing $I$ to the set of nodes $V_{\mathcal{P}}=\{x \in V :: \mathcal{P}(x)\}$.
A node that has received $I$ is termed \emph{covered}, and otherwise it is \emph{uncovered}.
The \Dis problem is solved at time slot $t_2\geq t_1$ if, for every node $v\in V_{\mathcal{P}}$, $v$ is covered by time slot $t_2$.
\end{definition}

The \Dis problem abstracts several common problems in distributed systems.
E.g. Broadcast, Multicast, Geocast, Routing etc., are all instances of this problem for a particular predicate $\mathcal{P}$. In order to prove lower bounds, we will use one of these instances: the Geocast problem. The predicate $\mathcal{P}$ for Geocast is $\mathcal{P}(x)=\mathrm{true}$ if and only if, at time $t_1$, $x$ is up and running, and it is located within a parametric distance $d>0$ (called \emph{eccentricity}) from the position of the source node at that time.

%\marginpar{\textcolor{red}{Text removed here}}
%Two models are considered in the Broadcast literature~\cite{baryehuda:broadcast,KP:broadcastupperbound,KP:adaptvsobliv,BDP:broadcast}. In the \emph{conditional} model, all nodes other than the source of broadcast cannot transmit until they receive the information to be disseminated for the first time. In the \emph{spontaneous} model, all nodes may transmit at any time. In this work, we generalize this classification to all \Dis problems
%%In particular, to prove our lower bound for adaptive protocols, we consider the conditional model, where nodes can not transmit before receiving the broadcast message for the first time.
%considering \emph{Conditional \Dis} protocols where a node does not transmit
%until it is covered (before receiving the \mess $I$ to be disseminated for the first time).

We explore how to solve this problem in a MANET where nodes are placed in a plane, in which they can move with bounded speed, and communication between nodes occurs over a collision-prone single channel. In this setup informed and uninformed nodes can be disconnected for some time (bounded by a parameter $\alpha$), but eventually some uninformed node must become neighbor of an informed node and remain so for some
time (bounded by a parameter $\beta$). In addition, nodes can start at different times, and they can crash and recover.
To solve the dissemination problem we consider three classes of randomized algorithms. In \emph{locally adaptive} randomized algorithms, the probability of transmission of a node in a step of an execution may depend on the communication history during such execution. On the contrary, in \emph{oblivious} randomized protocols, the probability of transmission of a node in a step of an execution depends only on predefined parameters.  Additionally we consider the class of \emph{fair} randomized protocols, in which at each step all nodes transmit with the same probability.

\label{section:relwork}

\parhead{Previous Work}
The literature on Information \Dis and related problems is vast and its complete review is beyond the scope of this article. Therefore, we focus in this section in the most relevant previous work with theoretical guarantees. The closest work to this paper is~\cite{FMMZ:geocast} where we studied \Dis under the same model but for deterministic protocols. One of the aims of this paper is to put those results together with the present work in perspective to study the impact that randomization and adaptiveness have on the time complexity of \Dis. 
%Further details are given in Section~\ref{sec:results}.

The \Dis problem abstracts several common problems in \RNs where some set of nodes hold some information that must be distributed to another set of nodes. Depending on the size of those sets, the problem receives different names, such as \emph{Broadcast}~\cite{baryehuda:broadcast,KM:broadcast} (one-to-all), \emph{k-Selection}~\cite{K:selection,CKR:m2m} (many-to-all), \emph{Gossiping}~\cite{FCM:gossiping, CGLP:gossiping} (all-to-all), or \emph{Multicast}~\cite{CKR:m2m,GKPX:m2m} (many-to-many). A geographic characterization of the set of receivers yields also specific problems, such as the Geocast problem~\cite{JC:geocastSurvey}.
%defined in Section~\ref{section:prelim}. 
Upper bounds for these problems may be used for \Dis depending on the receivers-set containment and mobility resilience, whereas a lower bound for any of them holds for the whole class as an existence lower bound.

Protocols that do not rely on knowledge of topology may allow node mobility. Topology-independent randomized protocols with theoretical guarantees have been studied, for Gossiping in~\cite{CGR:gossip} showing $O(n \log^4 n)$ expected time, and for Broadcast in~\cite{MRM:broadMANET} showing $O(n\log n)$ expected rounds. Both protocols require strong connectivity during all the execution and the latter additionally requires collision detection. 
The first randomized Broadcast protocol for Radio Networks was presented in~\cite{baryehuda:broadcast}. The protocol works in $O((D+\log n/\varepsilon)\log n)$ time with probability at least $1-\varepsilon$, where $D$ is the diameter of the network. More recently, the expectation to solve Broadcast was upper bounded by $O(D\log (n/D)+\log^2 n)$ in~\cite{CR:broadcastupperbound,KP:broadcastupperbound} for adaptive algorithms, and by $O(n \min\{D, \log n\})$ in~\cite{KP:adaptvsobliv} for oblivious protocols. None of these protocols is resilient to mobility. 
On the negative results side, a lower bound of $\Omega(D \log(n/D))$ was derived in~\cite{KM:broadcast} for the expected time for any randomized Broadcast protocol and a lower bound of $\Omega(\log^2 n)$ in~\cite{ABLP:broadcast}, matching together the upper bound in~\cite{CR:broadcastupperbound,KP:broadcastupperbound}. Later on, it was shown in~\cite{KP:adaptvsobliv} that, for every oblivious randomized Broadcast protocol, there exists a static network such that the protocol takes time in $\Omega(n)$ with probability at least $1/2$ to complete Broadcast. 
Given that a static network is a particular case of a MANET, these lower bounds apply to our setting. Nevertheless, we improve these bounds here by exploiting mobility.

Communication primitives have been studied for \emph{dynamic networks}, a suitable model for time-dependent topologies defined by a set of nodes and a sequence of edge-sets modeling the dynamic connectivity. If the topology defined by each edge set can be embedded in $\mathbb{R}^2$, a dynamic network is a suitable model of a MANET. 
Stochastic \Dis in dynamic networks has been studied.
The results in~\cite{BCF:parsimonious,CPMS:dynRNinfspread,CPMS:dynRNflooding} are probabilistic because the network is modeled as a stochastic time-dependent graph (edge-Markovian process), but the protocol is deterministic (flooding). 

In~\cite{CPMS:dynRN} the authors study randomized Broadcasting under two adversaries, one that for each step draws an Erdos-Renyi random graph from $\mathcal{G}_{n,p}$, and a so called \emph{worst-case adversary} that can make arbitrary changes to the topology for each step, as long as the adversary is \emph{meaningful}, i.e., at any time slot, it keeps at least one link on from an informed node to a non informed one. This latter adversary is equivalent to the adversary considered here for $\alpha=0$ and $\beta=1$, except for the speed limitations.
For the worst case adversary and for fair protocols, they show matching upper and lower bounds of $\Theta(n^2/\log n)$, with high probability and in expectation respectively. Here, we show the same lower bound for a weaker adversary and the same upper bound with high probability, both taking into account the time that the partition informed/covered is disconnected.

Deterministic upper and lower bounds for the problems above have been studied for MANETs~\cite{MCSPS:manetbroadcast,GS:multicast,PR:multicast,PSMCS:manetbroad,BDP:broadcast,DP:geomBroad}. 
Without relying on strong synchronization or stability assumptions, deterministic solutions for Geocast were recently proposed. In~\cite{FM:geocast}, the authors concentrate in the structure of the Geocast problem itself, whereas in~\cite{FMMZ:geocast}, tight lower bounds for different classes of protocols were shown taking into consideration the contention for the communication channel.

%\subsection{Our Results}
\label{sec:results}

\parhead{Our Results}
The aim of the present work is three-fold: \emph{(i)} to determine minimum values for parameters $\alpha$ and $\beta$ under which randomized protocols to disseminate information with big enough probability exist, \emph{(ii)} to study the time complexity in relation with the maximum speed of movement and the probability of failure, and \emph{(iii)} to put the results obtained here in perspective of our results in~\cite{FMMZ:geocast} in order to study the impact in time complexity of fundamental characteristics of dissemination protocols, such as determinism vs. randomization, and obliviousness vs. adaptiveness.

The results we obtain have been classified depending on the type of randomized protocol considered. First, for all fair protocols, an existential lower bound of $\Omega((n \log(1/\varepsilon))/\log n)$ on the time to increment the number of covered nodes in one unit with probability $1-\varepsilon$ has been derived. Then, for oblivious protocols and for \LA protocols 
an existential lower bound of $\Omega(n/\log n)$ on the time to increment the number of informed nodes, with high probability and in expectation respectively is given. 

A second collection of results lower bound the time required to complete \Dis.
In particular, it is shown that for the three types of protocols considered, fair, oblivious, and \LA, a similar lower bound of $\Omega(\alpha n +n^2/\log n)$ time steps exists, in order to solve the problem with probability $p\geq 2^{-n/2}$ for fair and oblivious protocols, and in expectation for the \LA class.
The proofs of these three results incur in increasing level of complexity, and all of them are presented because they include different constant factors and parameter values.

Finally, it is shown that a very simple fair oblivious protocol proposed in~\cite{CPMS:dynRN} can also be used to solve \Dis, guaranteeing termination in 
$O( \alpha n+(1+\frac{\alpha}{\beta})\frac{n^2}{\log n} )$
%$O( \alpha n(1+n/(\beta \log n)) + n^2/\log n )$ 
time, with probability $p\geq 1-e^{-(n-1)/4}$. Surprisingly, this bound holds for any value of $\alpha$, $\beta \geq 1$, arbitrarily large speed of movement, and arbitrary activation schedules. In this protocol, 
covered nodes always transmit with the same probability.
Observe that, when $\alpha/\beta=O(1)$, the time bound is asymptotically optimal for any of the protocol classes studied, 
which is rather surprising, given the simplicity of the protocol.

A summary of the bounds to solve \Dis presented in this paper is shown in Table~\ref{table:results}, together with the bounds for deterministic protocols obtained in~\cite{FMMZ:geocast}. The first important observation is that there is no gap between oblivious and \LA protocols. In fact, it seems that the oblivious class of protocols can be strengthen including adaptiveness, as long as local decisions are not positively correlated on previous events, as the techniques used to prove our lower bound only requires that.
The lower bounds derived match the upper bound shown using a fair oblivious protocol, for any $\alpha/\beta \in O(1)$. 
%(if $\beta \in \Omega(n/\log n)$). 
The second observation is that randomization reduces the time complexity of the problem in a linear factor in the oblivious case and in a logarithmic factor in the adaptive case (for reasonably small values of $\alpha$). It is important to note that all lower bounds in this paper have been proved without exploiting node failures or a non-simultaneous node activation, whereas non-simultaneous activation was crucial in showing a separation between oblivious and adaptive for deterministic protocols~\cite{FMMZ:geocast}.
On the other hand, it is fair to notice that the adaptive class of protocols considered in~\cite{FMMZ:geocast} is more general, and the constraints on speed assumed to prove the lower bounds are more restrictive.

\begin{table*}[tbp]\centering
\begin{small}
\begin{tabular}{l|l|c|c}
&&randomized&deterministic~\cite{FMMZ:geocast}\\
\hline
lower bounds&oblivious&
\rule{0pt}{3ex}$\Omega\left(\alpha n + n^2/\log n \right)$&$\Omega\left(\alpha n+n^3/\log n\right)$\\[.05in]
\cline{2-4}
&adaptive&\rule{0pt}{3ex}$\Omega\left(\alpha n +n^2/\log n\right)$&$\Omega(\alpha n+n^2)$\\[.05in]
\cline{2-4}
&fair&\rule{0pt}{3ex}$\Omega\left(\alpha n + n^2/\log n \right)$&--\\[.05in]
\hline
upper bounds&oblivious&\rule{0pt}{3ex}$O\left( \alpha n+\left(1+\alpha/\beta\right)n^2/\log n \right)$&$O(\alpha n+n^3\log n)$\\[.05in]
\cline{2-4}
&adaptive&--&\rule{0pt}{3ex}$O(\alpha n+n^2)$\\[.05in]
\cline{2-4}
&fair&\rule{0pt}{3ex}$O\left( \alpha n+\left(1+\alpha/\beta\right)n^2/\log n \right)$&--\\[.05in]
\multicolumn{4}{c}{}
\end{tabular}
\caption{Time complexity of Opportunistic Dissemination Information in \MANs.
All randomized lower bounds are to achieve success probability $p\geq 2^{-n/2}$, except the case of %general 
\LA which is the expected time to solve the problem. 
The randomized upper bound is with success probability $p\geq 1-e^{-(n-1)/4}$.}
\label{table:results}
\end{small}
\end{table*}

\parhead{Roadmap}
The rest of this paper is structured as follows. In Section~\ref{section:model} the model and definitions used are given. In
Sections~\ref{section:lb} and \ref{s-slb} we present our lower bounds, and the upper bound is proved in Section~\ref{section:ub}. The conclusions are presented in Section~\ref{s-concl}.
First, we state the following facts that will be used throughout the analysis.
\begin{align}
0\leq x<1 &\implies e^{-x/(1-x)} \leq 1-x \leq e^{-x}\label{eq-1}\\
0\leq x\leq 1/2 &\implies 4^{-x} \leq 1-x \label{eq-2}
\end{align}

Let $X_1,\dots,X_\ell$ be independent Poisson trials and $X=\sum_{i=1}^\ell X_i$. Then, the following Chernoff-Hoeffding bounds hold~\cite{book:mitzenmacher}.

For $0<\varphi<1$, 
\begin{align} 
Pr(X \leq (1-\varphi) E[X]) &\leq \left(\frac{e^{-\varphi}}{(1-\varphi)^{1-\varphi}}\right)^{E[X]}.\label{chernofftightbelow}\\
Pr(X \leq (1-\varphi) E[X]) &\leq e^{-\varphi^2E[X]/2}.\label{chernoffbelow}
\end{align} 

For $\varphi>0$, 
\begin{align} 
Pr(X \geq (1+\varphi) E[X]) &\leq \left(\frac{e^{\varphi}}{(1+\varphi)^{1+\varphi}}\right)^{E[X]}.\label{chernofftightabove}
\end{align} 

%For $0<\varphi\leq 1$, 
%\begin{align} 
%Pr(X \geq (1+\varphi) E[X]) &\leq e^{-\varphi^2E[X]/3}.\label{chernoffabove}
%\end{align} 

For $R\geq 6E[X]$, 
\begin{align} 
Pr(X \geq R) &\leq 2^{-R}.\label{chernofflooseabove}
\end{align}

\section{Model and Definitions}
%%%%%%%%%%%%%%%%%%%%%%%%%%%%%%%%%%%%%%%%%%%%

%\subsection{The Model}
\label{section:model}

\parhead{The Model}
We consider a MANET formed by a set $V$ of $n$ mobile nodes deployed in $\mathbb{R}^2$, where no pair of nodes can occupy the same point in the plane simultaneously.
It is assumed that each node has data-processing and radio-communication capabilities, and a unique identificator number (ID) in $[n]\triangleq\{1,\dots,n\}$. \footnote{The availability of an ID is not used in this paper and this assumption could be removed, but we include it for clarity.}

%\paragraph{Time:}
\emph{Time:}
Each node is equipped with a clock that ticks at the same uniform rate $\rho$; however, given the asynchronous activation, the clocks of different nodes may start at different times. A time interval of duration $1/\rho$ is long enough to transmit (resp. receive) the information to be disseminated. Computations in each node are assumed to take no time. Starting from a time instance used as reference, the global time is slotted as a sequence of time intervals or \emph{time slots} $1,2,\dots$, where slot $i>0$ corresponds to the time interval $[(i-1)/\rho,i/\rho)$. Without loss of generality~\cite{roberts:slottedaloha} all node's ticks are assumed to be in phase with this global tick.

%\paragraph{Node Activation:}
\emph{Node Activation:}
We say that a node is \emph{active} if it is powered up, and \emph{inactive} otherwise. It is assumed that, due to lack of power supply or other unwanted events that we call \emph{failures}, active nodes may become inactive. Likewise, due also to arbitrary events such as replenishing their batteries, nodes may be re-activated.
We call the temporal sequence of activation and failures of a node the \emph{activation schedule}.

We assume that a node is activated in the boundary between two consecutive time slots. If a node is activated between slots $t-1$ and $t$ we say that it is activated at slot $t$, and it is active in that slot.
Upon activation, a node immediately starts running from scratch an algorithm previously stored in its hardware (or firmware), but no other information or status is preserved while a node is inactive.
Consequently, it is possible that a covered node does not hold the \mess $I$, because it has been inactive after receiving it. To distinguish a covered node that does not hold the \mess from one that holds it, we introduce the following additional terminology: we say that a node $p$ is \emph{informed} at a given time $t$ if it holds the \mess $I$ at time $t$, otherwise $p$ is said to be \emph{uninformed}.

%\paragraph{Radio Communication:}
\emph{Radio Communication:}
Nodes communicate via a collision prone single radio channel. A node $v$ can receive a transmission of another node $u$ in time slot $t$ only if their distance is at most the \emph{range of transmission} $r$ during the whole slot $t$. The range of transmission is assumed to be the same for all nodes and all time slots. If two nodes $u$ and $v$ are separated by a distance at most $r$, we say that they are \emph{neighbors}.
In this paper, no collision detection mechanism is assumed, and a node cannot receive and transmit at the same time slot.
Therefore, an active node $u$ receives a transmission from a neighboring node $v$ at time slot $j$ if and only if $v$ is the only node in $u$'s neighborhood (including $u$ itself) transmitting at time slot $j$.
We say in this case that the transmission was \emph{successful}.
Also, a node cannot distinguish between a collision and no transmission. In general, we say that a node
$v \in V'$ \emph{transmits uniquely} among the nodes of set $V' \subseteq V$ in a slot $t$ if it is the only node in $V'$ that transmits in $t$.

%\paragraph{Link stability:}
\emph{Link stability:}
We assume that nodes may move on the plane. Thus, the topology of the network is time dependent. For simplicity, we assume that the topology only changes in the boundaries between time slots. Then, at time slot $t$ nodes $u$ and $v$ are connected by a link in the network topology if and only if they are neighbors during the whole slot $t$. An \emph{\OR} between two nodes $u$ and $v$ is a sequence of nodes $u=w_0, w_1, \ldots, w_k=v$ and a sequence of time slots 
$t^{(1)} < t^{(2)} < \cdots < t^{(k)}$ such that the network has a link between $w_{i-1}$ and $w_i$ at time slot $t^{(i)}$. 
We say that the network is \emph{potentially epidemic} if, after the initial time $t_1$, there is an \OR from the source $s$ to every node in $V_{\mathcal{P}}$.
Observe that in order to be able to solve an instance of \Dis, it is necessary that the network is potentially epidemic. 
%%
%\textcolor{blue}{shmuel: was this term defined?}
%\textcolor{green}{Miguel: I'm afraid not. Alessia: is there any reference we could use to avoid more definitions?}  
%\textcolor{red}{Anto: My suggestion is to remove the term, i.e.,  rephrase as "Observe that in order to be able to solve an instance of \Dis, it is necessary that after the initial time $t_1$, there is an \OR from the source $s$ to every node in $V_{\mathcal{P}}$".}
%\textcolor{green}{Miguel: epidemic is used later a few times, repeating this text would be cumbersome.}
%%
%I.e. after the initial time $t_1$, there is an \OR from the source $s$ to every node in $V_{\mathcal{P}}$.
However, worst-case adversarial choice of topologies for a dynamic network may preclude a randomized protocol from completing Broadcast with a desired probability as we show, even if connectivity is guaranteed. Note that Broadcast is an instance of \Dis, and that if there is connectivity then there are {\OR}s between all nodes. Thus, the property that the network is potentially epidemic as described is not sufficient to solve \Dis, and further limitations to the adversarial movement and activation schedule are in order.

While respecting a bound on the maximum speed $v_{\max}$, which is a parameter, the adversarial movement and activation schedule is limited by the following connectivity property:

\begin{definition}
\label{def:char2} Given a \MAN, an instance of the \Dis problem that starts at time $t_1$, and two integers $\alpha\geq 0$ and $\beta\geq 1$, the network is \emph{$(\alpha,\beta)$-connected} if, for every time slot $t \geq t_1$ at which the problem has not yet been solved, there is a
time slot $t'$ such that the following conditions hold:
\begin{itemize}
\item the intersection of time intervals $[t, t+\alpha]$ and $[t', t'+\beta)$ is not empty, and \item there is a pair of nodes $p$, $p'$, such that (a) at time $t'$, $p$ is informed and $p'$ is uncovered, and (b) $p$ and $p'$ are active and neighbors from time step $t'$ until $p'$ becomes covered, or until time step $t'+\beta-1$ (inclusive), whichever occurs first.\footnote{Observe that this model is slightly weaker than the one used in \cite{FMMZ:geocast}.}
\end{itemize}
\end{definition}

It is of the utmost importance to notice that $(\alpha,\beta)$-\-con\-nec\-tiv\-i\-ty is a characterization that applies to \emph{any} model of dynamic network, given that for any mobility and
%churn model
activation schedule, and any pair of nodes, there is a minimum time they are connected (even if that time is $1$) and a maximum time they are disconnected (even if that time is very large).
Thus, any dynamic network model used to study the \Dis problem has its own $\alpha$ and $\beta$ values.

Due to the same argument, $(\alpha,\beta)$-connectivity does not guarantee by itself that the network is epidemic (i.e. that the \mess is eventually disseminated); instead, an $(\alpha,\beta)$-connected network is only \emph{potentially} epidemic. Consider for instance the source node.
Thanks to the $(\alpha,\beta)$-\-con\-nec\-tiv\-i\-ty, at most every $\alpha$ slots, the source $s$ is connected to other nodes of the network for at least $\beta$ time slots. But, we have progress only if the protocol to solve \Dis is able to use the $\beta$ slots of connectivity to cover some uncovered node.
As a consequence of the above discussion, impossibility results only restrict $\beta$, whereas $\alpha$ only constrains the running time, as it is shown in this paper.

\emph{Adversary:}
We assume the presence of an adversary that controls each node activation schedule (including failures) and movement, restricted to the speed and connectivity constraints defined above.
This adversary is adaptive, in the sense that it makes decisions at the end of each step with access to all nodes' internal state, but without access to future random bits.
In order to obtain stronger results, we further restrict the adversary while proving lower bounds. More precisely, all our lower-bounds hold, even if all nodes are activated simultaneously and never fail. Such an assumption is not a minor restriction for the adversary, since it implicitly provides extra resources that protocols may exploit, such as a global time.
In fact, assuming that nodes could be activated at different times was instrumental in the study of deterministic protocols
\cite{FMMZ:geocast}.
However, as we show, in this case it does not help.

%\subsection{Protocols for \Dis}
\label{sec:protocols}

\parhead{Protocols for \Dis}
Recall that in our model the time is slotted. In the following, we use the term \emph{time slot} to refer to a global time of reference, and the term \emph{time step} to refer to the time reference local to a node. 

As customary in the \RNs literature~\cite{KP:adaptvsobliv,CGLP:gossiping}, we classify randomized protocols as \emph{oblivious} or \emph{adaptive}.
In oblivious protocols, the probability of transmission of a node in a step of an execution depends only on predefined parameters. E.g. the node ID~\footnote{Even if nodes do not have ID's a different program may be stored in each node before deployment.} or the number of steps the node has been active. In other words, the protocol is oblivious of communication history. Formally,

\begin{definition}
A randomized protocol for Conditional Dissemination in a MANET formed by a set of nodes $V$ is called \emph{oblivious} if it can be modeled by a set $\Pi=\{\pi_i|i\in V\}$ of transmission-probability sequences $\pi_i=\langle\pi_{i1},\pi_{i2},\dots\rangle$ such that, for any execution,
a covered\footnote{Recall that a node is covered if it has received the \mess $I$.} node $i$ transmits independently with probability $\pi_{it}$ in the $t$-th step after activation.
\end{definition}

In adaptive algorithms, the probability of transmission of a node in a step of an execution may additionally depend on the communication history during such execution. E.g., information contained in the transmissions received or the number of transmission trials. 
%\textcolor{green}{Due to pragmatic reasons (feasibility of our analysis) 
In the present work, adaptiveness is restricted to \emph{any} local information. I.e., nodes may adapt arbitrarily as long as environment (e.g. other nodes, channel, etc.) information is not used.
It is worth to note that the techniques used in our lower bound proofs can be also used for a more general class of adaptive protocols where the probability functions in Definition~\ref{d-adaptive} are correlated, as long as such correlation is not positive. Although smaller, we focus in the class defined below for clarity. 
The study of more general adaptive protocols is left for future work.
The definition follows.

\begin{definition}
\label{d-adaptive}
Let $\mathcal{P}$ be a randomized protocol for Conditional Dissemination in a MANET formed by a set of nodes $V$.
%Let $m(i,t)$ be a set of pairs $\langle t',m\rangle$ such that message $m$ was received by node $i$ in step $t'\leq t$ while executing $\mathcal{P}$.
Let $s(i,t)$ be a sequence of $t$ symbols from the alphabet $\{T,R\}$ describing the transmission/reception schedule of node $i$ during the first $t$ steps executing $\mathcal{P}$. 
Then, $\mathcal{P}$ is called \emph{\LA} if it can be modeled by a set $\Pi=\{\pi_i|i\in V\}$ of transmission-probability functions 
%$\pi_i(t,m(i,t-1), s(i,t-1))$ 
$\pi_i(t, s(i,t-1))$
such that, 
a covered node $i$ transmits with probability $\pi_i(t, s(i,t-1))$ in the $t$-th step after activation.
\end{definition}

%It is worth to note that the techniques used in our lower bound proofs can be also used for a more general class of adaptive protocols where the probability functions in Definition~\ref{d-adaptive} are correlated, as long as such correlation is not positive. Although smaller, we focus in the class defined above for clarity.

%Oblivious algorithms are of interest in \RNs because they are simple and easy to implement, as opposed to adaptive algorithms that are usually more intricate and may involve computational overhead, but they are usually more efficient. However, to the best of our knowledge, adaptive protocols for \Dis in MANETs that are theoretically faster are not known, and the lower bound shown here for a particular class of adaptive protocols is asymptotically tight with respect to the oblivious upper bound.

We further characterize oblivious and adaptive algorithms considering the class of \emph{fair} protocols. A protocol is fair if within the same time slot, with respect to the global time, all active nodes transmit with the same probability. Formally,

\begin{definition}
Let $\mathcal{P}$ be a randomized protocol for Conditional Dissemination in a MANET formed by a set of nodes $V$.
For any execution $\chi$ of protocol $\mathcal{P}$ on $V$, let $\rho_{it}(\chi)$ be the probability of transmission of node $i\in V$ at time slot $t$ during the execution $\chi$ of $\mathcal{P}$.
Then, $\mathcal{P}$ is called \emph{fair} if $\rho_{it}(\chi)=\rho_{jt}(\chi)$ for all $\chi$, $t$, and covered nodes $i,j\in V$.
\end{definition}

Notice that fairness is an orthogonal property to obliviousness since both, oblivious and adaptive protocols, may be fair. For the former to be fair it implies that all nodes use the same probability of transmission in all time steps when node activation is adversarial.

%%%%%%%%%%%%%%%%%%%%%%%%%%%%%%%%%%%%%%%%%%%%%%%%%%%%%%%%

%\section{Lower Bounds}

\section{Link Stability Bounds}
\label{section:lb}

Link stability assumptions are crucial in solving any \Dis problem. In our model, such characteristic is parameterized with $\beta$. In this section, we relate link stability to the worst-case running time and the error probability~\footnote{We use ``error'' from the perspective of Monte Carlo algorithms.}   of Geocast protocols.
Notably, exploiting fairness, it is possible to quantify such relation no matter how strong are the stability guarantees (i.e. for any $\beta>0$), which we do in the first theorem.

%%%%%%%%%%%%%%%%%%%%%%%%%%%%%%%%%%%%%%%%%%%%%%%%%%
\begin{figure}[htbp]
\begin{center}
\psfrag{<r}{$\leq r$}
\psfrag{<eps}{$\leq \xi$}
\psfrag{>r<eps}{$r<\delta\leq r+\xi$}
\psfrag{A}{$A$}
\psfrag{B}{$B$}
\psfrag{C}{$B'$}
\includegraphics[width=0.7\textwidth]{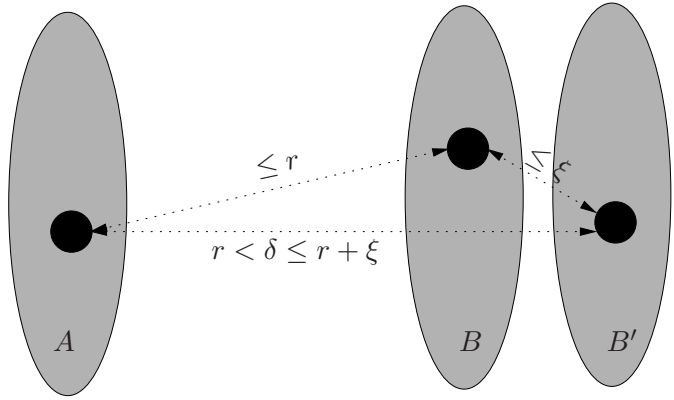}
\caption{Illustration of Theorems~\ref{thm:betafair},~\ref{thm:betachernoff}, and~\ref{thm:lb-adaptive}.}
\label{fig:imposs}
\end{center}
\end{figure}
%%%%%%%%%%%%%%%%%%%%%%%%%%%%%%%%%%%%%%%%%%%%%%%%%%

\begin{theorem}
\label{thm:betafair}
For
any $V_{max}>0$,
$d>r$,
$0<\varepsilon<1$,
$\alpha\geq 0$,
$\beta>0$,
any $k,c \in \mathbb{Z}^+$, such that $45\leq k<n$ and $c>0$,
and for any fair randomized protocol $\mathcal{P}$ for Geocast, 
%\textcolor{blue}{can delete 'as defined in section 2'}  as defined in Section~\ref{sec:protocols},
there exists an $(\alpha,\beta)$-connected MANET formed by a set $V$ of $n$ nodes running $\mathcal{P}$ such that,
if $T \leq k\log_4(1/\varepsilon)/(4\log k)$,
within $c T$ steps after $k$ nodes are covered,
no new node is covered with probability at least $\varepsilon^c$,
even if all nodes are activated simultaneously and do not fail.
\end{theorem}

\begin{proof}
Consider three sets of nodes $A$, $B$, and $B'$ deployed in the plane, each set deployed in an area of size $\xi$ arbitrarily small, such that $0<\xi<r$ and $d\geq r+\xi$. (See Figure~\ref{fig:imposs}.)% in the Appendix.)
The invariant in this configuration is that nodes in each set form a clique, every node in $A$ is placed within distance $r$ from every node in $B$, every node in $B$ is placed at most at distance $\xi$ from every node in $B'$, and every node in $A$ is placed at some distance $r<\delta\leq r+\xi$ from every node in $B'$.
Also, $\xi$ is set appropriately so that a node can move $\xi$ distance in one time slot without exceeding $V_{max}$.

Initially, the adversary places $k$ nodes in the set $B'$ including the source,
the remaining $n-k$ nodes in set $A$, and the set $B$ is left initially empty.
At the beginning of the first time slot all nodes are activated.
%Anto
Let $t$ be the first time slot when all the nodes in $B'$ have been informed, i.e., the information has been delivered
for the first time to some node in $B'$ within time slot $t$. (If this event never happens
the claim of the theorem holds trivially.)
According to Definition~\ref{def:char2}, $(\alpha,\beta)$-connectivity has been preserved until time slot $t$.
The adversary does not move any nodes until then.
Given that $d\geq r+\xi$, the nodes in $A$ must become informed to solve the problem.

After time slot $t$, the adversary moves the nodes according to protocol $\mathcal{P}$ as follows.
The adversary places any node $y\in B'\cup B$ in $B$ forever.
This node preserves $(\alpha,\beta)$-connectivity as long as the nodes in set $A$ are not informed.
For each time slot $t'> t$ where nodes transmit with probability at least $4\log k / k$, the adversary moves all nodes in $B'$ to $B$.
At the end of time slot $t'$ the adversary moves all nodes in $B$ but $y$ back to $B'$, and the procedure is repeated.

We show now that within the first $T\leq \ln(1/\varepsilon) k/(2\ln k)$ steps the nodes in $A$ are not informed with probability at least $\varepsilon$.
Given that the protocol is fair, in a given time slot all nodes use the same probability of transmission.
Let $\pi_{t'}$ be the probability of transmission used in step $t'$.
The probability of not achieving a successful transmission in a step $t'$ is
%\begin{align*}
$Pr_{fail} = 1 - \sum_{i\in B} \pi_{t'} \prod_{j\in B, j\neq i} (1-\pi_{t'})
=  1 - |B| \pi_{t'} (1-\pi_{t'})^{|B|-1}$.
%\end{align*}
We consider two cases
depending on whether $\pi_{t'}$ is bigger or smaller than $4\log k / k$. For both cases it can be shown that $Pr_{fail} \geq 4^{-4\log k/k}$. The details are omitted in this extended abstract for brevity.
%
%\emph{Case 1:} $\pi_{t'}\geq 4\log k / k$. Then,
%\begin{align*}
%Pr_{fail} &= 1 - |B| \pi_{t'} (1-\pi_{t'})^{|B|-1}\\
%&\geq 1 - k (1-4\log k /k)^{k-1}; \textrm{ since $(4\log k)/k<1$ for $k>16$,}\\
%&\geq 1 -  \frac{k}{e^{4(k-1)\log k /k}}, \textrm{ from Eq~(\ref{eq-1}); since $k/(e^{4(k-1)\log k /k})\leq 1/2$ for $k>1$,}\\
%&\geq 4^{-k/e^{4(k-1)\log k /k}}, \textrm{ from Eq~(\ref{eq-2});}\\
%&\geq 4^{-4\log k/k}
%\end{align*}
%
%%The latter inequality holds because
%%\begin{align*}
%%4^{-k/e^{4(k-1)\log k /k}} &\geq 4^{-4\log k/k}\\
%%4^{-k/e^{4(k-1)\log k /k}} &\geq 4^{-1/k}\\
%%k^2  &\leq  e^{4(k-1)\log k /k}\\
%%2k\ln k  &\leq  4(k-1)\log k\\
%%k\ln 2  &\leq  2(k-1), \textrm{ for any $k>1$.}
%%\end{align*}
%
%
%\emph{Case 2:} $\pi_{t'}< 4\log k / k$. Then,
%\begin{align*}
%Pr_{fail} &= 1 - |B| \pi_{t'} (1-\pi_{t'})^{|B|-1}\\
%&> 1 - 4\log k/k;\textrm{ since $4\log k /k \leq 1/2$ for $k\geq 45$,}\\
%&\geq 4^{-4\log k/k}, \textrm{ from Eq~(\ref{eq-2}).}
%\end{align*}
%
%
%
Then, the probability of failing to inform the nodes in set $A$ within the interval $[t+1,t+T]$ is
%\begin{align*}
$Pr_{fail} (T \textrm{ steps})
\geq 4^{-4T\log k/k}
\geq \varepsilon$, if
$T \leq k\log_4(1/\varepsilon)/(4\log k)$.
%\end{align*}
Conditioned on this event, the same analysis can be applied to the subsequent interval of $T$ steps, and inductively to each subsequent interval of $T$ steps. %Thus, the claim follows.
%\qed
\end{proof}

%%%%%%%%%%%%%%%%%%%%%%%%%%%%%OBLIVIOUS LINK STABILITY

%The following theorem pertains to oblivious protocols but possibly not fair. 
For oblivious protocols, possibly not fair, it is shown in the following theorem that,
%It is shown that, 
if the link stability guarantees are not strong enough, there exists some configuration where new nodes are not covered with positive probability.
The proof %, left to the appendix for brevity, 
uses the %same 
adversarial configuration of Theorem~\ref{thm:betafair}. However, given that nodes running oblivious protocols may use different probabilities of transmission in a given time slot, in order to preserve the $(\alpha,\beta)$-property the node $y$ has to be chosen more carefully.

\begin{theorem}
\label{thm:betachernoff}
For
any $V_{max}>0$,
$d>r$,
$\alpha\geq 0$,
$\beta>0$,
any $k,c \in \mathbb{Z}^+$, such that $e^3\leq k<n$ and $c>0$,
and any oblivious randomized protocol $\mathcal{P}$ for Geocast,  as defined in Section~\ref{sec:protocols},
there exists an $(\alpha,\beta)$-connected MANET formed by a set $V$ of $n$ nodes running $\mathcal{P}$ such that
if $\beta<k/(2(1+\ln k))$,
within $c\beta$ steps after $k$ nodes are covered,
no new node is covered with probability at least $(2e^e)^{-c}$,
even if all nodes are activated simultaneously and do not fail.
\end{theorem}

\begin{proof}
%The proof shows an adversarial configuration and movement of nodes that proves the claimed lower bound. The configuration, initial deployment, and movement of nodes are the same of Theorem~\ref{thm:betafair}, except that now the adversary has to chose more carefully the node $y$ to preserve $(\alpha,\beta)$-connectivity.
%Also, the analysis requires to take into account all steps where some node $y$ is left alone in $B$ together.
%The complete proof is left to the Appendix for brevity.

Consider the same configuration, initial deployment, and worst-case assumptions used in Theorem~\ref{thm:betafair} up to step $t$.
(Recall that the set $B'$ has $k$ nodes and that at time slot $t$ all of them have been informed.)
After time slot $t$, the adversary moves the nodes according to protocol $\Pi$ as follows.
The adversary moves a node $y$ from $B'$ to $B$ to satisfy $(\alpha,\beta)$-connectivity.
(The way that node $y$ is chosen will be described in the analysis.)
The adversary places $y$ in $B$ for all time slots in the interval $[t+1,t+\beta]$, hence satisfying $(\alpha,\beta)$-connectivity.
Additionally, for each time slot $t'\in[t+1,t+\beta]$ where $\sum_{j\in B\cup B'}\pi_{jt'}\geq 1+\ln k$, the adversary moves all nodes from $B'$ to $B$.
At the end of each time slot $t'$ the adversary moves all nodes in $B$ but $y$ back to $B'$.
At the end of time slot $t+\beta$, the adversary moves $y$ back to $B'$, and the procedure is repeated.

We show now that if $\beta<k/(2(1+\ln k))$, within the first $\beta$ steps the nodes in $A$ are not informed with probability at least $1/(2e^e)$. Consider any step $t'\in[t+1,t+\beta]$ such that

{\bf Case 1:} $\sum_{j\in B\cup B'}\pi_{jt'}\geq 1+\ln k$.

Letting $X(j)$ be a random variable indicating whether node $j$ transmits or not at time step $t'$, and $X=\sum_{j\in B\cup B'} X(j)$ a random variable indicating the number of transmissions at time step $t'$, by linearity of expectation, the expected number of transmissions at step $t'$ is $E[X] = \sum_{j\in B\cup B'} E[X(j)] = \sum_{j\in B\cup B'} \pi_{jt'} \geq 1+\ln k$. Given that the random variables $X(j)$ are independent,
setting $\varphi=1-1/E[X]<1$ in Inequality~(\ref{chernofftightbelow}), because $E[X]\geq 1+\ln k$ and  $k>1$, we have
\begin{align*}
Pr (X\leq 1) &\leq \left(\frac{e^{-1+1/E[X]}}{(1/E[X])^{1/E[X]}}\right)^{E[X]}\\
&= \frac{eE[X]}{e^{E[X]}}, \textrm{ letting $E[X]=x+\ln k$ for some $x\geq 1$,}\\
&= \frac{e(x+\ln k)}{k e^x}, \textrm{ given that $x\geq 1$,}\\
&\leq \frac{e\ln k}{k}.
\end{align*}

Then, if $T_1$ is the set of steps $t'\in [t+1,t+\beta]$ such that $\sum_{j\in B\cup B'}\pi_{jt'} \geq 1+\ln k$.
\begin{align*}
Pr_{fail} (\textrm{in }T_1)
&\geq \left(1-\frac{e\ln k}{k}\right)^\beta, \textrm{ given that $k >e$,}\\
&\geq \exp\left(-\frac{e\beta\ln k}{k-e\ln k}\right), \textrm{ from Eq~\ref{eq-1},}\\
&> \frac{1}{e^e}.
\end{align*}

The latter inequality holds for $\beta<k/(2(1+\ln k))$ and $k\geq e^3$.

{\bf Case 2:} $\sum_{j\in B\cup B'}\pi_{jt'}< 1+\ln k$.

Let $T_2$ be the set of steps $t'\in [t+1,t+\beta]$ such that $\sum_{j\in B\cup B'}\pi_{jt'}< 1+\ln k$. Given that $|T_2|\leq\beta$, the sum of probabilities over all nodes and all those steps is
$$\sum_{j\in B\cup B', t''\in T_2}\pi_{jt''} < \beta(1+\ln k).$$
Given that there are $k$ (informed) nodes in $B\cup B'$, there exists a node $y$ such that, the sum of all probabilities of transmission of $y$ over all those steps is
$$\exists y\in B\cup B' : \sum_{t''\in T_2}\pi_{yt''} <\beta(1+\ln k) / k.$$
The adversary chooses such node as the node $y$ to place in $B$ during all steps in $[t+1,t+\beta]$.
Given that $\beta< k/(2(1+\ln k))$ we know that $ \sum_{t''\in T_2}\pi_{yt''}< 1/2$ and consequently $\forall t''\in T_2 : \pi_{yt''}< 1/2$.
Then,
\begin{align*}
Pr_{fail} (\textrm{in }T_2)
&= \prod_{t''\in T_2}(1-\pi_{yt''})\nonumber\\
&\geq 4^{-\sum_{t''\in T_2}\pi_{yt''}}, \textrm{ from Eq~(\ref{eq-2}),} \nonumber\\
&> \frac{1}{2}, \textrm{ for $\beta<\frac{k}{2(1+\ln k)}$}.
\end{align*}

For each of the above cases, the lower bound shown also holds if the number of steps of that case is smaller. Hence, the probability of failing to inform nodes in $A$ within the interval $[t+1,t+\beta]$ is at least $1/(2e^e)$.

Conditioned on the event analyzed above, the same argument can be applied to the subsequent interval of $\beta$ steps, and inductively to each subsequent interval of $\beta$ steps. Thus, the claim follows.

\end{proof}%\qed

%%%%%%%%%%%%%%%%%%%%%%%ADAPTIVE LINK STABILITY

We move now to \LA protocols possibly not fair as defined in Section~\ref{sec:protocols}. Recall that the adversary is adaptive, making decisions at the end of each step with access to all the nodes' internal state, but without access to their future random bits.
For this class, it is shown that if $\beta$ is not large enough, there exists some configuration where it is not expected to inform a new node within the first $\beta$ steps.
The proof %, left to the appendix for brevity, 
uses the same adversarial configuration of Theorems~\ref{thm:betafair} and~\ref{thm:betachernoff}. However, given that nodes running \LA protocols may change the probabilities of transmission according to local history, the node $y$ that preserves $(\alpha,\beta)$-property is chosen conveniently to show that the choice is correct with big enough probability.

\begin{theorem}
\label{thm:lb-adaptive}
For any \LA randomized protocol $\mathcal{P}$ for Geocast,  as defined in Section~\ref{sec:protocols},
such that uninformed nodes never transmit,
any $V_{max}>0$,
$d>r$,
$\alpha\geq 0$,
$\beta>0$,
and any $k\in \mathbb{Z}^+$, such that $(2/(1-1/e))^{\xi/e}< k<n$, $\xi\triangleq 2/(1-1/e)^2$,
there exists an $(\alpha,\beta)$-connected MANET formed by a set $V$ of $n$ nodes such that
if $\beta< k/(2e\Gamma)$, for $\Gamma\triangleq \xi\ln\delta$, $\delta\triangleq\beta^2k^{e/\xi}$, within $\beta$ steps after $k$ nodes were covered,
in expectation no new node is covered,
even if all nodes were activated simultaneously and do not fail.
\end{theorem}

\begin{proof}
Consider the same configuration, initial deployment, and worst-case assumptions used in Theorems~\ref{thm:betafair}  and~\ref{thm:betachernoff} up to step $t$.
(Recall that the set $B'$ has $k$ nodes and that at time slot $t$ all of them have been informed.)
After time slot $t$, the adversary moves the nodes as in Theorem~\ref{thm:betachernoff}, but now the procedure to choose node $y$ is more elaborate.
The adaptive nature of $\mathcal{P}$ does not preclude the adversary from knowing when to move all nodes to $B$, since that may be done on a step-by-step basis as in Theorem~\ref{thm:betachernoff}. However, given that the choice of node $y$ has to be done in advance for a whole interval of $\beta$ steps, adaptiveness is a major obstacle for the adversary.
Before showing how to decide which is the node $y$, we define some necessary random variables and we compute some necessary bounds as follows.

Let $T$ be the set of steps in the interval $[t+1,t+\beta]$.
For any $t'\in T$, let $X_{jt'}$ be a random variable indicating whether node $j$ transmits or not at time step $t'$, and $X_{t'}=\sum_{j\in B\cup B'} X_{jt'}$ a random variable indicating the number of transmissions at time step $t'$, by linearity of expectation, the expected number of transmissions at step $t'$ is $E[X_{t'}] = \sum_{j\in B\cup B'} E[X_{jt'}] = \sum_{j\in B\cup B'}\pi_j(t',s(j,t'-1))$.
Given that the $X_{jt'}$ are independent Poisson trials,
we can use Chernoff-Hoefding concentration bounds as follows.
If $E[X_{t'}]\geq \Gamma$ at $t'$, using Inequality~(\ref{chernoffbelow}),
$$Pr\left(X_{t'}\leq \Gamma/e\right) \leq Pr\left(X_{t'}\leq E[X_{t'}]/e\right)\leq\exp(-E[X_{t'}]/\xi)\leq 1/\delta.$$
Otherwise, if $E[X_{t'}]< \Gamma$ at $t'$, using Inequality~(\ref{chernofftightabove}),
$$Pr\left(X_{t'}> e\Gamma\right) \leq (E[X_{t'}]/\Gamma)^{e\Gamma}e^{-E[X_{t'}]}\leq 1/\delta.$$
The later inequality holds because $(E[X_{t'}]/\Gamma)^{e\Gamma}e^{\Gamma/\xi}\leq e^{E[X_{t'}]}$, which can be verified considering both subcases of $E[X_{t'}]\lessgtr\Gamma/\xi$.

In order to choose node $y$, the adversary carries out the following computation.
For $t+1$, the adversary knows all transmission probabilities that will be used.
So, for each $j\in B\cup B'$, it computes $E[X_{jt+1}]$ without conditioning on previous events, i.e. $E[X_{jt+1}]=\pi_j(t+1,s(j,t))$.
However, for $t+2$, the adversary has to condition the computation on the outcome of the previous step.
For each $j\in B\cup B'$, if $E[X_{t+1}]\geq\Gamma$  (resp. $E[X_{t+1}]<\Gamma$) the adversary computes $E[X_{jt+2}|X_{t+1}>\Gamma/e]$ (resp. $E[X_{jt+2}|X_{t+1}\leq e\Gamma]$).
Notice that such computation is feasible since the adversary has access to all nodes' internal state (including the protocol itself) and consequently it can easily compute all possible outcomes under the assumption that $X_{t+1}>\Gamma/e$ (resp. $X_{t+1}\leq e\Gamma$).
Inductively, the same argument can be applied for each of the successive steps up to $t+\beta$.
Let these expectations computed conditioning on previous events be denoted as $E'[\cdot]$.

For each step, the event on which the computation is conditioned upon occurs with probability at least $1-1/\delta$ as shown above. Thus, for each $j\in B\cup B'$ and for all $t'\in T$, the values $E'[X_{jt'}]$ computed occur with probability at least $1-\beta/\delta$ (using the union bound over the $\beta$ steps). Based on this computation, consider the set of steps $T_2=\{t'\in T|E'[X_{t'}]<\Gamma\}$. As shown above, we have that $\sum_{t'\in T_2}E'[X_{t'}]<\beta e\Gamma$ with the same probability. Thus, there is a node $j\in B\cup B'$ for which $\sum_{t'\in T_2}E'[X_{jt'}]<\beta e\Gamma/k$ with probability at least $1-\beta/\delta$, and the adversary choose precisely this node $j$ as the node $y$ to preserve $(\alpha,\beta)$-connectivity. Then, given that $\beta< k/(2e\Gamma)$ we have that, with the same probability, $\sum_{t'\in T_2}E'[X_{jt'}]< 1/2$.

Consider now the slots in $T-T_2$. As shown above, with probability at least $1-\beta/\delta$, for each $t'\in T-T_2$ it holds that $E'[X_{t'}] \geq \Gamma/e$. Thus, if the adversary choose the slots where $E[X_{t'}] \geq \Gamma/e$ as the slots of high contention to move all nodes to $B$ (which can be done on a step-by-step basis), with the same probability the choice of node $y$ is correct (i.e., node $y$ transmits with sufficiently low probability when it is left alone in $B$).
Given that the random variables  $X_{jt'}$ are independent, and that $0<1-1/E[X_{t'}]<1$ because $k>e$, setting $\varphi=1-1/E[X_{t'}]$ in Inequality~(\ref{chernofftightbelow}), the probability that there is a successful transmission in slot $t'$ is at most
\begin{align*}
Pr (X_{t'}\leq 1) &\leq \left(\frac{e^{-1+1/E[X_{t'}]}}{(1/E[X_{t'}])^{1/E[X_{t'}]}}\right)^{E[X_{t'}]}\\
&= \frac{eE[X_{t'}]}{e^{E[X_{t'}]}}\\
&\leq \frac{\Gamma}{e^{\Gamma/e}}, \textrm{ because $\Gamma/e\geq 1$.}
\end{align*}

Thus, with probability at least $1-\beta/\delta$, the expected number of successful transmissions within slots in $T-T_2$ is at most $\beta\Gamma/e^{\Gamma/e}$. Replacing $1\leq\beta<k/(2e\Gamma)$ and $\Gamma=\xi\ln\delta$, it is at most $k/(2e\delta^{\xi/e}) = 1/(2e\beta^{2\xi}) \leq 1/2e$. Finally, it remains to consider the expected number of transmissions that occur with probability at most $\beta/\delta$, which is at most $\beta$ since there are $\beta$ slots in $T$. Therefore, overall, if $\beta< k/(2e\Gamma)$ the expected number of successful transmissions in $T$ is less than
\begin{align*}
\left(1-\frac{\beta}{\delta}\right)\left(\frac{1}{2}+\frac{1}{2e}\right)+\frac{\beta}{\delta}\beta
&=\left(1-\frac{1}{\beta k^{e/\xi}}\right)\left(\frac{1}{2}+\frac{1}{2e}\right)+\frac{1}{k^{e/\xi}}\\
&<\frac{1}{2}+\frac{1}{2e}+\frac{1}{k^{e/\xi}}\\
&<1, \textrm{ for $k>(2/(1-1/e))^{\xi/e}$.}
\end{align*}
\end{proof}%\qed

\section{\Dis Lower Bounds}
\label{s-slb}

In this section, we show that there exist an instance of \Dis, namely Geocast, for which adversarial configurations of nodes that require a minimum number of steps exist.
%%%%%%%%%%%%%%%%%%%%%%% GEOCAST FAIR

%%%%%%%%%%%%%%%%%%%%%%%%%%%%%%%%%%%%%%%%%%%%%%%%%% 
\begin{figure}[htbp]
\begin{center}
\psfrag{A}{$A$}
\psfrag{B}{$B$}
\psfrag{B'}{$B'$}
\psfrag{C}{$C$}
\psfrag{X}{$x$}
\psfrag{Y}{$y$}
\psfrag{r}{$\leq r$}
\psfrag{eps}{$\leq \varepsilon$}
\psfrag{d}{$r<\cdot\leq r+\varepsilon$}
\psfrag{u}{$u$}
\psfrag{v}{$v$}
\psfrag{w}{$w$}
\psfrag{s}{$s$}
\includegraphics[width=0.7\textwidth]{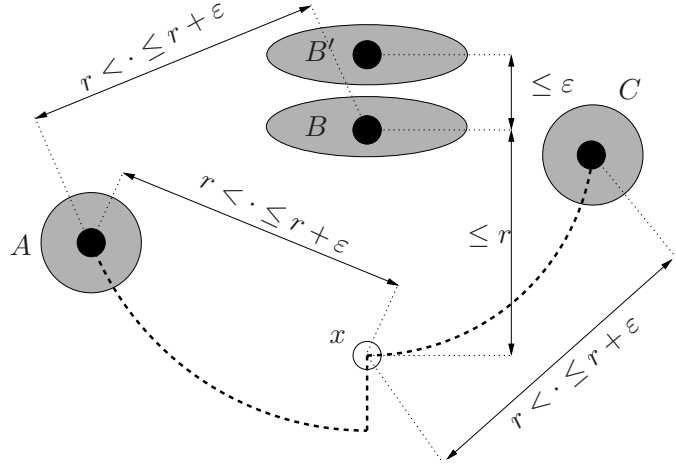}
\caption{Illustration of Theorems~\ref{thm:geocastfair},~\ref{thm:geocastobliv}, and~\ref{thm:geocastadaptive}. Distances invariant. A small empty circle depicts a point in the plane. A small black circle depicts a node. A big empty area depicts an empty set. A big shaded area depicts a non-empty set.}
\label{fig:geocastlb}
\end{center}
\end{figure}
%%%%%%%%%%%%%%%%%%%%%%%%%%%%%%%%%%%%%%%%%%%%%%%%%% 

\begin{theorem}
\label{thm:geocastfair}
For any $n> 24$, $d>r$, $\alpha\geq 0$, $\beta>0$,
$V_{max}> \pi r / (6 \alpha)$,
and any \emph{fair} randomized Geocast protocol $\mathcal{P}$, as defined in Section~\ref{sec:protocols},
there exists an $(\alpha,\beta)$-connected MANET of $n$ nodes
for which, in order to solve the problem with probability at least $2^{-n/2}$,
$\Pi$ takes at least $\alpha n/2 +n^2/(96\ln (n/2))$ time slots.
\end{theorem}

\begin{proof}
We prove the claim for fair protocols where the probability of transmission of a node is independent among time slots. For non-independent fair protocols, the same techniques can be used to prove the same bound, with some more detail in the analysis. %\textcolor{blue}{i find the next two sentences confusing, and am not sure what they intend to say.  may be they can be erased? } However, it is not clear for us that an instance of such class of fair protocols even exists. Therefore, we omit those details for clarity of presentation.

The following adversarial configuration and movement of nodes show the claimed lower bound. Consider four sets of nodes $A$, $B$, $B'$, and $C$, each deployed in an area of size $\varepsilon$ arbitrarily small, such that $0<\varepsilon<r$ and $d\geq r+\varepsilon$, and a point $x$ all placed in the configuration depicted in Figure~\ref{fig:geocastlb}. % in the Appendix.
The invariant in these sets is the following:
all nodes in each set form a clique;
each node in $A$ is placed at some distance $>r$ and $\leq r+\varepsilon$ from the point $x$ and each node in $B$;
each node in $B$ is placed within distance $r$ of the point $x$;
each node in $B'$ is placed within distance $\varepsilon$ of each node in $B$ and $> r$ from the point $x$ and each node in $A$;
each node in $C$ is placed at some distance $>r$ and $\leq r+\varepsilon$ from the point $x$ and distance $>r$ from each node in $A$.

At the beginning of the first time slot, the adversary places $n/2$ nodes, including the source node $s$, in set $B'$, the remaining $n/2$ nodes in the set $A$, and starts up all nodes. (For clarity, assume that $n$ is even.)
The other two sets are initially empty.
Given that $d\geq r+\varepsilon$, all nodes must be covered to solve the problem.
Also, $\varepsilon$ is set appropriately so that a node can be moved $\varepsilon$ distance in one time slot without exceeding $V_{max}$, and so that a node can be moved from set $A$ to point $x$ through
the curved part of the dotted line, of length less than $\pi (r+\varepsilon)/6$, in $\alpha$ time slots without exceeding $V_{max}$. (To see why the length bound is that, it is useful to notice that the distance between each pair of singular points along the circular dotted line is upper bounded by $(r+\varepsilon)/2$.)

Let $t$ be the first time slot when the source is scheduled to transmit. Adversarially, let $t$ be the first time slot when the source is informed. Nodes stay in the positions described until $t$ and, consequently, all the other $n/2 - 1$ nodes in set $B'$ receive it. Hence, until time slot $t$, $(\alpha,\beta)$-connectivity has been preserved. Starting at time slot $t+1$, the adversary moves the nodes so that only one new node at a time becomes informed.

First, we describe broadly the movements and later we give the details.
Some of the nodes in $B'$ are moved back and forth to $B$ to produce contention.
Nodes in $A$ are moved one by one following the dotted lines in two phases, first up to point $x$, and afterwards to the set $C$.
While a node moves from point $x$ to $C$ a new node moves from $A$ to $x$.
The procedure is repeated until all nodes in $A$ are covered.

The movement of each node $u$ moved from $A$ to $C$ is carried out in two phases of $\alpha$ time slots each separated by an interlude as follows.
\begin{itemize}
\item \emph{Phase 1.}
During the first $\alpha-1$ time slots, $u$ is moved from $A$ towards the point $x$ maintaining a distance $>r$ and $\leq r+\varepsilon$ with respect to every node in $B$.
Nodes in $B'$ stay static during this interval.
Given that only nodes in $B'$ are informed and the distance between them and $u$ is bigger than $r$, $u$ does not become covered during this interval.
In the $\alpha$-th time slot of this phase, $u$ is moved to the point $x$ and any node $y\in B'$ is moved to $B$, preserving $(\alpha,\beta)$-connectivity until $u$ is informed.
Upon reaching point $x$, $u$ and all the nodes in $A$ and $C$ remain static until Phase 2.

\item \emph{Interlude.}
During this interval, nodes in $B'$ are moved back and forth to $B$ according to protocol $\Pi$ to produce contention as follows.
For each time slot where nodes transmit with probability at least $8\ln (n/2) / n$, the adversary moves all nodes in $B'$ to $B$. At the end of the time slot the adversary moves all nodes in $B-\{y\}$ back to $B'$, and the procedure is repeated until $u$ is covered when the interlude ends.
At the end of such time slot all nodes in $B$ are moved to $B'$.

\item \emph{Phase 2.}
During the following $\alpha$ slots, $u$ is moved towards the set $C$ while a new node $v$ is moved from the set $A$ towards the point $x$. I.e., Phase 2 of node $u$ is executed concurrently with Phase 1 of node $v$ (hence, nodes in $B'$ stay static during this interval). The nodes $u$ and $v$ are moved in such a way that they maintain a distance $>r$ between them.
At the end of this phase $u$ is placed in set $C$ and stays static forever.
At this point node $v$ has reached point $x$, but $u$ can not cover $v$ because all nodes in $C$ are at distance greater than $r$ from $x$.

\end{itemize}

The movement detailed above is produced for each node initially in $A$, overlapping the phases as described, until all nodes have became covered.
In each phase of at least $\alpha$ time slots every node is moved a distance at most $\pi (r+\varepsilon)/6+\varepsilon$. Thus, setting $\varepsilon$ appropriately, the adversarial movement described does not violate $V_{max}$.

We prove now the time bound.
For any time slot $t$ in the interludes, the probability of covering the node in $x$ is
%\begin{align*}
$P = \sum_{i\in B} \pi_t \prod_{j\in B, j\neq i} (1-\pi_t)
=  |B| \pi_t (1-\pi_t)^{|B|-1}$.
%\end{align*}
For any $t$ when $\pi_t< 8\ln (n/2) / n$, we have $P < 8\ln (n/2)/n$ because in this case the adversary puts just a single node in $B$, which is $y$.
On the other hand, for any $t$ when $\pi_t\geq 8\ln (n/2) / n$, we also have
%\begin{align*}
%P &\leq \frac{n}{2} \left(1-\frac{8\ln (n/2)}{n}\right)^{n/2-1}; \textrm{ since $\frac{8\ln (n/2)}{n}<1$ for $n>24$,}\\
%&\leq \frac{n}{2e^{8(n/2-1)\ln (n/2) /n}}, \textrm{ from Eq~\ref{eq-1},}\\
%&\leq \frac{8\ln (n/2)}{n}
%\end{align*}
$P \leq (n/2) \left(1- 8\ln (n/2)/n\right)^{n/2-1}$ since $8\ln (n/2)/n<1$ for $n>24$.
Using Eq~\ref{eq-1}, % in the Appendix, 
we have $P \leq n/(2e^{8(n/2-1)\ln (n/2) /n}) \leq 8\ln (n/2)/n$.
%The latter inequality holds because
%\begin{align*}
%\frac{n^2}{16\ln (n/2)}  &\leq  e^{8(n/2-1)\ln (n/2) /n}\\
%\frac{n^2}{4}  &\leq  e^{8(n/2-1)\ln (n/2) /n}\\
%n\ln (n/2)  &\leq  4(n/2-1)\ln (n/2)\\
%n  &\leq  4(n/2-1)
%\end{align*}

Let $X$ be a random variable denoting the number of successful transmissions along $T=n^2/(96\ln(n/2))$ interlude steps.
The expected number of successful transmissions is $E[X]=TP\leq n/12$.
Given that $X$ is the sum of independent Poisson trials, using 
Chernoff bounds,
%Inequality~(\ref{chernofflooseabove}),
%If steps were not independent, and there were some coin tossing to decide, we would need to explain how the adversary computes the probability of transmission from the probability of those coins (Bayes). We omit it because it seems that such protocol does not exist and it would only mess the presentation.
For $n/2\geq 6E[X]$,
%\begin{align*}
$Pr(X \geq n/2) \leq 2^{-n/2}$.
%\end{align*}
We conclude that $T$ interlude steps are necessary to cover all nodes in $A$ with probability at least $2^{-n/2}$. On the other hand, Phase 1 of all nodes in $A$ adds $\alpha n/2$ steps to the overall time. Thus, the claim follows.
%\qed
\end{proof}

%%%%%%%%%%%%%%%%%%%%%%% GEOCAST OBLIVIOUS

We move now to prove an existential lower bound for \Dis oblivious protocols, possibly not fair. 
The proof %, left to the appendix for brevity, 
uses the same adversarial configuration of Theorem~\ref{thm:geocastfair}, but given that nodes running oblivious protocols may use different probabilities of transmission in a given time slot, the node that preserves the $(\alpha,\beta)$-property for each newly informed node has to be chosen more carefully as in Theorem~\ref{thm:betachernoff}.

\begin{theorem}
\label{thm:geocastobliv}
For any $n>3$, $d>r$, $\alpha\geq0$, $\beta>0$,
$V_{max}> \pi r / (6 \alpha)$,
and any \emph{oblivious} randomized Geocast protocol $\Pi$,  as defined in Section~\ref{sec:protocols},
there exists an $(\alpha,\beta)$-connected MANET of $n$ nodes for which,
$\Pi$ takes at least $\alpha n/2+n^2/(48e\ln(n/2))$ time slots
in order to solve the problem with probability at least $2^{-n/2}$.
\end{theorem}

\begin{proof}
%The proof shows an adversarial configuration and movement of nodes that proves the claimed lower bound. The configuration, initial deployment, and movement of nodes are the same of Theorem~\ref{thm:geocastfair}, except that now the adversary has to chose more carefully the node $y$ to preserve $(\alpha,\beta)$-connectivity.
%Also the analysis, requires to take into account all steps where some node $y$ is left alone in $B$ together.
%The complete proof is left to the Appendix for brevity.

The following adversarial configuration and movement of nodes shows the claimed lower bound. Consider the same configuration, initial deployment, and movement of nodes of Theorem~\ref{thm:geocastfair}, except for the following. (Refer to Figure~\ref{fig:geocastlb}.)

\begin{itemize}
\item \emph{Phase 1.}
The node $y\in B'$ to be moved to $B$ in the $\alpha$-th time slot of this phase is carefully chosen as described in the analysis.

\item \emph{Interlude.}
During this interval, nodes in $B'$ are moved back and forth to $B$ according to protocol $\Pi$ to produce contention as follows.
For each interlude time slot $t$ where $\sum_{j\in B\cup B'}\pi_{jt}\geq 1+\ln (n/2)$, the adversary moves the nodes in $B'\cup B-\{y\}$ to $B$.
At the end of each time slot $t$ the adversary moves all nodes in $B\cup B'-\{y\}$ back to $B'$.
The procedure is repeated until $u$ is informed or $n/(24e\ln (n/2))$ time slots have elapsed, whatever happens first.
At the end of such time slot all nodes in $B$ are moved to $B'$ and the interlude ends.

\item \emph{Phase 2.}
The movement of $u$ from point $x$ to the set $C$ is produced maintaining a distance at most $r$ with respect to $B\cup B'$ so that, if $u$ left point $x$ before being covered, $(\alpha,\beta)$-connectivity is still preserved.

\end{itemize}

We prove now the time bound.
We consider first only interlude steps.
Let $I$ denote the set of steps of interlude $I$.
For any step $t\in I$, we consider two cases.

{\bf Case 1:} $\sum_{j\in B\cup B'}\pi_{jt}\geq 1+\ln (n/2)$.

Letting $X(j)$ be a random variable indicating whether node $j$ transmits or not at time step $t'$, and $X=\sum_{j\in B\cup B'} X(j)$ a random variable indicating the number of transmissions at time step $t$, by linearity of expectation, the expected number of transmissions at step $t$ is $E[X] = \sum_{j\in B\cup B'} E[X(j)] = \sum_{j\in B\cup B'} \pi_{jt} \geq 1+\ln (n/2)$. Given that the random variables $X(j)$ are independent,
%using the following Chernoff-Hoeffding bound~\cite{book:mitzenmacher}.
%For $0<\varphi<1$,
%\begin{align*}
%Pr(X \leq (1-\varphi) E[X]) \leq \left(\frac{e^{-\varphi}}{(1-\varphi)^{1-\varphi}}\right)^{E[X]}
%\end{align*}
setting $\varphi=1-1/E[X]<1$ in Inequality~(\ref{chernofftightbelow}), because $E[X]\geq 1+\ln (n/2)$ and  $n>2$ the probability of covering the node in $x$ is
\begin{align*}
P\leq Pr (X\leq 1) &\leq \left(\frac{e^{-1+1/E[X]}}{(1/E[X])^{1/E[X]}}\right)^{E[X]}\\
&= \frac{eE[X]}{e^{E[X]}}, \textrm{ letting $E[X]=x+\ln (n/2)$ for some $x\geq 1$,}\\
&= \frac{2e(x+\ln (n/2))}{n e^x}, \textrm{ given that $x\geq 1$,}\\
&\leq \frac{2e\ln (n/2)}{n}.
\end{align*}

Let now $X_1$ be a random variable denoting the number of successful transmissions along $n^2/(48e\ln(n/2))$ interlude steps of case 1. The expected number of transmissions is $E[X_1]=Pn^2/(48e\ln(n/2))\leq n/24$.

{\bf Case 2:} $\sum_{j\in B\cup B'}\pi_{jt}< 1+\ln (n/2)$.

Let $T_2\subseteq I$ be the set of steps $t\in I$ such that $\sum_{j\in B\cup B'}\pi_{jt}< 1+\ln (n/2)$. I.e., the set of steps in case 2.
Given that $|T_2|\leq|I|\leq n/(24e\ln (n/2)) \leq n/(24(1+\ln (n/2)))$ (the latter for $n\geq 2e^{1/(e-1)}$), the sum of probabilities over all nodes and all those steps is
$$\sum_{j\in B\cup B', t'\in T_2}\pi_{jt'} < n/24.$$
Given that there are $n/2$ (informed) nodes in $B\cup B'$, there exists a node $y$ such that, the sum of all probabilities of transmission of $y$ over all those steps is
$$\exists y\in B\cup B' : \sum_{t'\in T_2}\pi_{yt'} <1/12.$$
The adversary chooses such node as the node $y$ to place in $B$ during all steps in the interlude $I$.
The adversary may choose a different node $y$ for each interlude, but we know that the upper bound holds for all of them.
Thus, given that there are exactly $n/2$ interludes in total, we know that the sum of the probability of covering the node in $x$ in a step of case 2 is at most $n/24$. Notice that such summation is the expectation of a random variable $X_2$ denoting the number of successful transmissions along all interlude steps of case 2. I.e. $E[X_2]\leq n/24$.

We consider now a new random variable $Y=X_1+X_2$ that denotes the number of successful transmissions along all steps of all interludes. We know that $E[Y]=E[X_1]+E[X_2]\leq n/12$.
Then, given that $Y$ is the sum of independent Poisson trials, the probability of solving the problem is, using Inequality~(\ref{chernofflooseabove}),
%the following Chernoff-Hoeffding bound~\cite{book:mitzenmacher}.
for $n/2\geq 6E[Y]$,
\begin{align*}
Pr(Y \geq n/2) &\leq 2^{-n/2}.
\end{align*}

We conclude that, in order to cover all nodes in $A$ with probability at least $2^{-n/2}$, the above overall time of $n^2/(48e\ln(n/2))$ for interlude steps is needed. On the other hand, Phase 1 of all nodes in $A$ adds $\alpha n/2$ steps to the overall time. Thus, the claim follows.

\end{proof}%\qed

%%%%%%%%%%%%%%%%%%%%%%% GEOCAST ADAPTIVE

The following theorem for \LA protocols, can be proved as a straightforward repeated application of Theorem~\ref{thm:lb-adaptive} to the configuration and movement of nodes described in Theorem~\ref{thm:geocastobliv}, changing the minimum contention under which informed nodes are moved appropriately. The complete proof is omitted for brevity.

\begin{theorem}
\label{thm:geocastadaptive}
For any
$n>17$, %$>2(2/(1-1/e))^{2/(e(1-1/e)^2)}$ and $>3$
$d>r$, $\alpha\geq 0$, $\beta>0$,
$V_{max}> \pi r / (6 \alpha)$,
and any \emph{\LA} randomized Geocast protocol $\Pi$,  as defined in Section~\ref{sec:protocols},
there exists an $(\alpha,\beta)$-connected MANET of $n$ nodes for which,
$\Pi$ takes on expectation at least $\alpha n/2+e^2(e+1)^2n^2/(2(e-1)^2\ln(n/2))$ time slots
in order to solve the problem.
\end{theorem}

%%%%%%%%%%%%%%%%%%%%%%%%%%%%%%%%%%%%%%%%%%%%%%%%%%%%%%%%%

\section{Upper Bound}
\label{section:ub}

The \Dis protocol analyzed in this section is fair and oblivious. The protocol was studied in~\cite{CPMS:dynRN} for Broadcast in dynamic networks.
% (modulo minor details in the probability of transmission introduced here for simplicity in the analysis). 
We show here that the same protocol can be used for \Dis in MANETs. In this protocol every informed node transmits the information $I$ at each time step with probability $p=\ln n/n$.
%The proof is left to the appendix for brevity.

%\begin{algorithm}[t]
%\label{alg}
%\caption{Fair-oblivious \Dis algorithm for each node. $I$ is the information to be disseminated.}
%\SetKwFor{Upon}{upon}{do}{endupon}
%\dontprintsemicolon
%\Upon{receiving $I$}{
%\For{each time step}{
%transmit $I$ with probability $p=\frac{\ln n}{n}$\;
%}}
%\end{algorithm}

\begin{theorem}
\label{thm:ubfair}
For any $(\alpha,\beta)$-connected MANET where $\beta\geq 1$, $n>2$, and any $V_{max}>0$, the fair-oblivious randomized protocol described
%detailed in Algorithm~\ref{alg} 
solves \Dis in time
$O( \alpha n+(1+\frac{\alpha}{\beta})\frac{n^2}{\log n})$
%$O(\alpha n(1+n/(\beta \log n))+n^2/\log n)$ 
steps with probability at least $1-e^{-(n-1)/4}$.
\end{theorem}

\begin{proof}
As a worst case, we assume that all nodes in the network must be covered, and $V_{max}$ is arbitrarily big.
The time slot when the information is assigned to the source node is $t_1$.
Let any time step $t> t_1$ in which an active uncovered node is connected to at least one active informed node be called a \emph{good} step.
From the proof of Theorem 3.1 in~\cite{CPMS:dynRN}, it is known that the probability that an uncovered node is informed in a good step is at least
$p/2$. Let $X_i$ be a random variable indicating that in the $i$th good step an uncovered node is informed (assuming an infinite supply of uncovered nodes for simplicity). Applying a Chernoff-Hoeffding bound, the probability that after
$S=4n(n-1)/\ln n$ good steps some node (of the $n-1$ that have to be covered after $t_1$) still has to be covered can be bounded as follows.
Let $X=\sum_{i=1}^S X_i$, $\mu=E[X]\geq Sp/2=2(n-1)$, and $\delta=1/2$, using Inequality~(\ref{chernoffbelow}),
%\begin{align*}
$Pr(X \leq n-1) = Pr(X \leq (1-\delta) \mu) \leq e^{-(n-1)/4}$.
%\end{align*}
From Definition~\ref{def:char2}, after any sequence of at most $\alpha$ bad steps there must be a sequence of good steps that lasts $\beta$ steps unless the uncovered node is informed before that. Hence, the total number of steps to have
$S$ good steps is at most
$O(\alpha(n+S/\beta)+S) = O( \alpha n+(1+\frac{\alpha}{\beta})\frac{n^2}{\log n})$.
\end{proof}

\section{Conclusions and Open Problems}
\label{s-concl}

Lower bounds on link stability and the time complexity of disseminating information in \MANs were studied in this work for fair, oblivious, and \LA protocols.
The results obtained show that, with respect to obliviousness, adaptiveness does not help if based on localized information. 
Furthermore, the techniques used in our lower bound can be also used for a more general class of adaptive protocols as long as the probability functions in Definition~\ref{d-adaptive} are not positively correlated among nodes.
The particular version of the problem studied was Conditional \Dis. We conjecture that for the spontaneous version, where nodes transmit before receiving the Information to disseminate, the lower bounds can be improved, although only up to constants.
A comparison of the bounds obtained with previous deterministic results~\cite{FMMZ:geocast} show that randomization reduces the complexity in a linear factor for oblivious protocols and at least (it might be even bigger for arbitrary adaptiveness) in a logarithmic factor for adaptive ones. 
Given that adversarial node-activation and node-failures were not used to prove lower bounds (as opposed to~\cite{FMMZ:geocast} where it was crucial), further exploration of arbitrarily-adaptive randomized protocols exploiting them is promising, and it is left for future work.

\section{Acknowledgements}
\label{section:ack}
We thank Seth Gilbert for triggering the development of this work asking a question at DISC 2010.

%\newpage
\bibliographystyle{abbrv}
\bibliography{./Comprehensive_2010}

%\appendix
%\section{Appendix}
%\input{appendix-anto}
%\input{appendix}
\end{document}